\documentclass[aps,pre,twocolumn,amsmath,amssymb,reprint]{revtex4-1}

\usepackage{graphicx}
\usepackage{dcolumn}
\usepackage{bm}

\usepackage[utf8]{inputenc}
\usepackage[T1]{fontenc}
\usepackage{mathptmx}
\usepackage{etoolbox}
\usepackage[english]{babel}

\makeatletter
\def\@email#1#2{%
 \endgroup
 \patchcmd{\titleblock@produce}
  {\frontmatter@RRAPformat}
  {\frontmatter@RRAPformat{\produce@RRAP{*#1\href{mailto:#2}{#2}}}\frontmatter@RRAPformat}
  {}{}
}%
\makeatother

\begin{document}

\preprint{AIP/123-QED}

\title[A trick of the tail: how electrostatics helps a DNA repair enzyme to localize on nucleosomes]{A trick of the tail: how electrostatics helps a DNA repair enzyme to localize on
nucleosomes}

\author{Safwen Ghediri$^{1,2}$, Guillaume Brysbaert$^{3}$, 
Fabrizio Cleri$^{1,2}$, Ralf Blossey$^{3}$}
\affiliation{$^{1}$ University of Lille, Institut d'Electronique, de Micro\'electronique et de Nanotechnologie (IEMN CNRS UMR8520)} 
\affiliation{$^{2}$ D\'epartement de Physique, 
59652 Villeneuve d'Ascq, France}
\affiliation{$^{3}$ University of Lille, CNRS, Unit\'e de Glycobiologie Structurale et Fonctionnelle (UGSF), UMR8576, 59000 Lille, France}

\date{\today}

\begin{abstract}
Electrostatic interactions are key to the recognition processes of proteins and DNA and have been previously documented for the action of repair enzymes. Uracil-DNA glycosylase (UDG) is the first in a sequence of enzymes that act in the base-excision repair process (BER) and whose task is the extraction of uracil bases from nuclear DNA. The question of how the molecule targets uracil bases in chromatin, in particular in the condensed protein-DNA complexes of nucleosomes, has only recently become a subject of detailed studies. Here we show that the presence of an arginine anchor motif on the N-terminal tail of UDG can favor its localization on nucleosomes by binding to their acidic patches on their top and bottom surfaces via electrostatic interactions. We argue that this mechanism can play a key role in the detection of uracil defects in nucleosomal DNA. 
\end{abstract}

\maketitle

\section{\label{sec:Introduction} Introduction}

The presence of wrong bases in DNA, as e.g. the RNA-base uracil in place of thymine, initiates the DNA repair process of base excision (BER) that leads to the replacement of uracil by the correct base. The BER process consists of a sequence of repair steps performed by different enzymes; its initial step is the recognition of the wrong base and its excision. Uracil base excision is performed by Uracil-DNA glycosylases (UDG), a class of a homologous enzymes.\cite{krokan2013} While the atomistic details of the excision process are well understood from the known crystal structures of UDG bound to DNA since a long time,\cite{slupphaug1996} how UDG actually finds its target -- notably in the context of chromatin 
-- is still a matter of debate.\cite{li2019} A typical scenario, for which in vitro experimental evidence exists, is that the enzyme diffuses along the DNA and searches for the `wrong' bases.\cite{kong2020,bigman2023} Attempts to approximately quantify this process for the in vivo situation, however, show that the detection rates 
that can be expected from this process are considerably less than what is needed and observed in the BER process in vivo; see, e.g., the detailed discussion in the Supplementary Material of Ref.\cite{saravanan2025} The quantitative estimate given there would amount to scanning time of days in the chromatin compartment, which is biologically unrealistic.
\\
{\it Target search.} Recently, we have found computational evidence that, in a model scenario closer to cellular reality, dynamic distortions of DNA in nucleosomal arrays suffice to destabilize the double-helix in such a way that a deformation-induced opening of the DNA allows for a flipping of the base pairs out of the double helix.\cite{cleri2023} These processes render the flipped bases accessible to enzymes located in the vicinity of the DNA strand in the nucleus. A subsequent study of the recognition of such flipped-out uracil base pairs established that for mechanically exposed bases, the recognition of uracil by UDG can be as good as in the formation of a full complex in solution, provided the base is fully flipped.\cite{saravanan2025} Thus, the presence of UDG enzymes near mechanically-induced open flipped out base pairs becomes a relevant process in a direct comparison with the spontaneous opening of the base pairs by thermal fluctuations. 
\\
{\it The role of electrostatics in target search.} Alongside mechanical forces acting on chromatin that allow for an induced flipping of uracil and can hence render it more easily detectable to UDG molecules in the vicinity, electrostatic effects may also play a role in the recognition process, and this in two possible ways. First, electrostatic effects may promote the diffusion of proteins along the DNA as pointed out in a theoretical study.\cite{dahirel2009} The electrostatic properties of hUNG have been studied with kinetic assays.\cite{cravens2014}  
On the other hand, the computation of the electrostatic attraction between UDG and its target base has been studied with numerical solutions of the Poisson-Boltzmann equation \cite{xie2021} using the
Delphi package.\cite{li2012} The key results of that paper are that the recognition of the flipped uracil is determined by a larger surface area of the UDG enzyme than just the active site, which in fact is negatively charged as the surface of the DNA. A quantitative estimate of the range of the electrostatic force of attraction between UDG and the DNA duplex yields a value of about 4 nm, and thus points to a relevant recognition mechanism based on electrostatic interactions.
\\
{\it Target identification in chromatin/nucleosomes.} The basic building block of chromatin, the nucleosome, presents a significant hindrance for the access by UDG, as has been shown several years ago.\cite{ye2012,biechele2022} Recent kinetic assays support this picture, and establish that the accessibility of uracils in nucleosomes, even in unfavorable positions close to the histone core, is favoured by less strongly positioning sequences than the commonly used Widom 601 sequence.\cite{barbosa2025} Recent computational studies by us have uncovered details of the interaction of UDG with uracils in differently accessible positions in nucleosomes based on strongly positioned sequences.\cite{ghediri2025,ghediri2025-2}
\\
{\it The N-terminal tail of UDG.} Most experimental and simulation studies of UDG have ignored the N-terminal tail of the molecule. This is understandable from the point of crystallography, and indeed all available results point to only 
minor effects on the direct recognition between UDG and uracil. There are, however, other effects in which this 50-residue long tail can play an important role, and one of these is indeed the diffusion of proteins along DNA, see the discussion of such findings in.\cite{vuzman2009,rodriguez2019,perkins2021} 
\\
{\it Objective of the present study.} In this paper we point to yet another effect of electrostatics due to the properties of the N-terminal tail of UDG with particular relevance for the tracing of uracil defects in nucleosomal DNA. The UDG N-terminal tail contains a nuclear localization signal (NLS), i.e. a short amino-acid sequence that signals the molecule transport apparatus to place where the molecule has to be transported to, in our case the nucleus.\cite{otterlei1998}
This sequence is rich in arginines, i.e. positively charged. \cite{perkins2021} The RKR-motif has been implied to have regulatory roles in the context of enzymatic activity 
\cite{wang2012} and, very recently, in the binding of a
chromatin remodeler domain to DNA.\cite{yadav2025}
Here we show that this localization motif is, at the same time, an arginine anchor motif that can bind to the acid patch of the nucleosome.\cite{mcginty2021,zhang2021} Hence, our results suggest that the interaction of the arginine anchor motif on the UDG tail and the acid patch of the nucleosome promote the presence of UDG near nucleosomal DNA and therefore can help increasing the uracil removal rate.
\\
{\it Conception of our study.}
Our study elucidates three aspects of the interaction between the full UDG molecule and the nucleosome. Firstly,
we take a look a the charge distribution of the UDG tail
and compute its electrostatic profile with the help of publicly available Poisson-Boltzmann servers. We also built a coarse-grained polymer model to simulate the tail dynamics and 
electrostatic recognition.
Secondly, we show by a homology analysis that the arginine-rich motif on the tail is not only a nuclear localization signal, but
bears similarity to known binding sequences of transcription factors to nucleosomes. Finally, we perform a structural modeling study of the binding of i) tail peptides containing the NLS sequence, and ii) the full UDG enzyme to a nucleosome. We end the paper with a conclusion of our findings and a discussion on their possible consequences for the action of UDG on nucleosomes.

\section{\label{sec: Results} Results}

\subsection{\label{sec: UDG electrostatics} Electrostatics of the UDG tail}

The structure of human uracil-DNA glycosylase was obtained from available crystallographic structures of the catalytic domain. As the N-terminal region of UDG is not resolved in these structures, which reflects
its high conformational flexibility, in order to obtain a structural representation of this missing segment, the full-length sequence of human uracil-DNA glycosylase was retrieved from UniProt (UniProt entry P13051-1, https://www.uniprot.org/uniprotkb/P13051). The predicted three-dimensional structure including the N-terminal tail was obtained from the AlphaFold Protein Structure Database
(https://alphafold.ebi.ac.uk/entry/P13051).\cite{jumper2021} The AlphaFold prediction provides a complete structural model of the protein, including the otherwise unresolved N-terminal residues. The resulting UDG model therefore includes both the catalytic core and the predicted N-terminal extension containing the arginine anchor (RKR) motif considered in this study.

\begin{figure}[h!]
\centering
\includegraphics[width=1.0\linewidth]{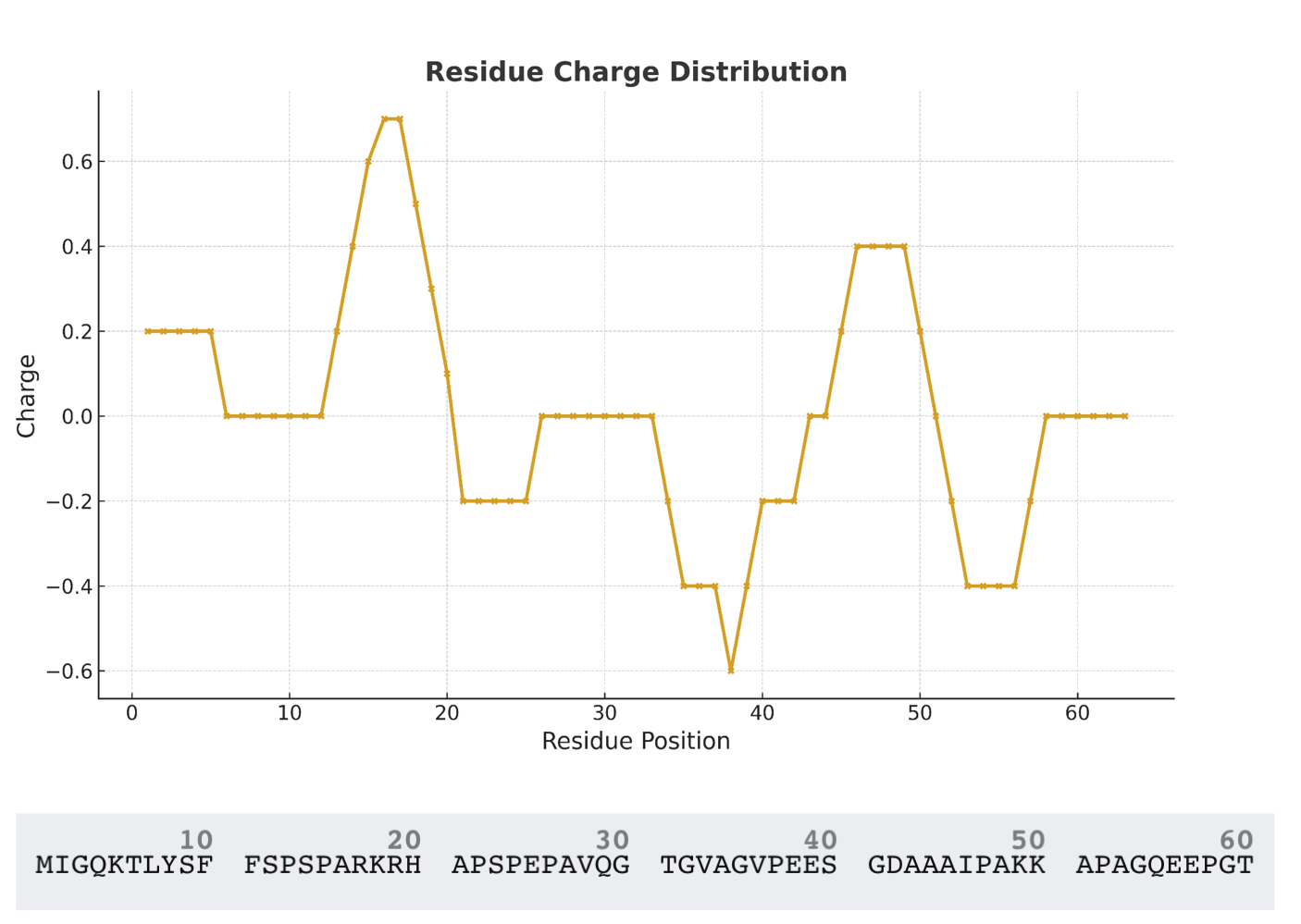}
\caption{Charge distribution of residues 1-50 along the UDG tail. Computed with EMBOSS.}
\label{fig:tail_charge}
\end{figure}

The first step in our study is the characterization of the electrostatic properties of the UDG tail. A simple
analysis of the charge composition of the tail obtained with the EMBOSS suite \cite{rice2000} \url{https://emboss.sourceforge.net/} is displayed in Figure  \ref{fig:tail_charge} of the complete protein hUNG, i.e. the charge distribution along the end of the N-terminal tail (residues 1-50). The presence of the positively charged motif --RKR-- on residues 17-19 is clearly noticeable in the charge distribution. We will therefore specifically target this region in the following analysis. 

\begin{figure}[h!]
\centering
\includegraphics[width=1.0\linewidth]{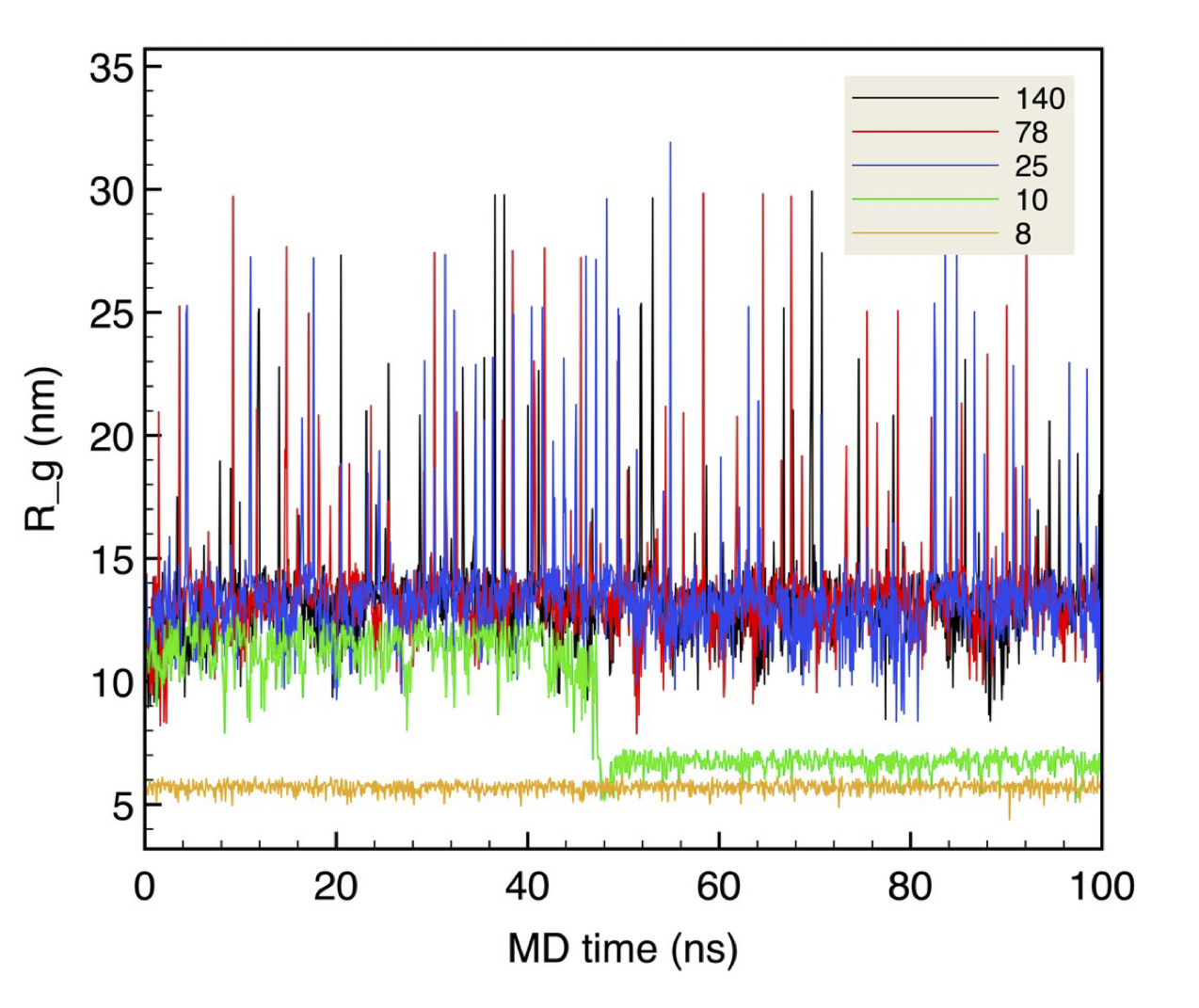}
\includegraphics[width=1.0\linewidth]{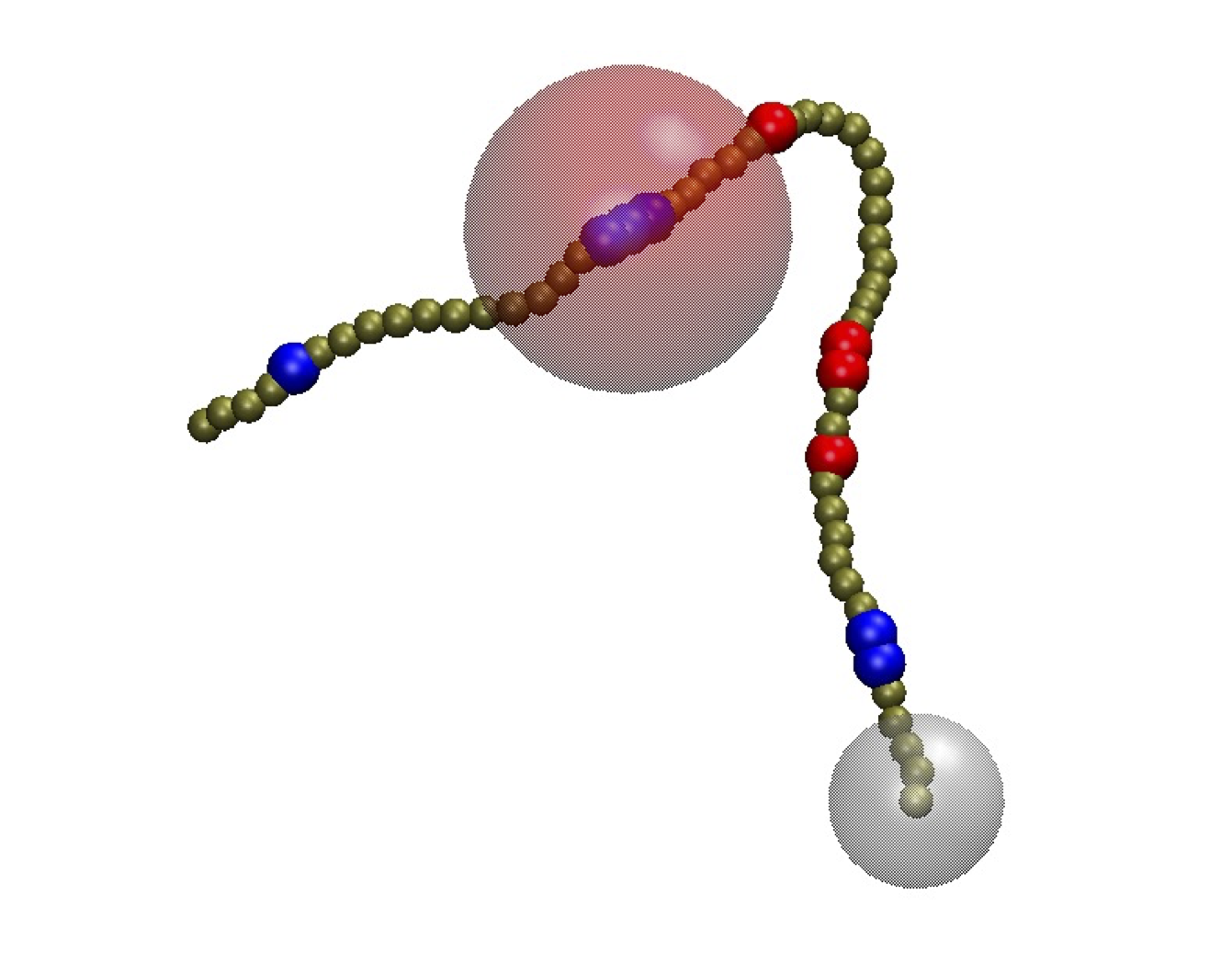}
\caption{a) Radius of gyration of the polymer tail model as a function of simulation time, for different values of the dielectric permittivity $\varepsilon$. b) The coarse-grained polymer model for the interaction of the tail fixed to a grey sphere (representing UDG body), with a negatively
charged red sphere (nucleosome) Blue/red residues in the chain are positively/negatively charged.}
\label{fig:figfc1}
\end{figure}

We tested the dielectric properties of the UDG tail in the nucleoplasm environment by building a simple coarse-grained model of a tethered polymer with the corresponding charge distribution. Each residue is represented by a sphere of 1 nm, linked into a chain by harmonic bond-stretching and bond-bending forces, and interacting with a soft-core wall potential to avoid self-crossing. Partially screened charges of –0.5 e and +0.5 e are set at Asp-Glu, and Arg-Lys residues, respectively. An implicit solvent is simulated by a distance-dependent dielectric (DDD), with adjustable permittivity $\varepsilon$. By running 500 ns MD simulations of these  configurations, we measured the radius of gyration $R_g$ 
, see Figure \ref{fig:figfc1} a). We observe that at 
$\varepsilon > 25 $ the chain is fully solvated and remains extended, while at lower permittivity it tends to fold onto itself, by crossing its negative and positive charged spots. Since the nucleoplasm is considered a dielectric medium with permittivity typically comprised between values of about 40-150, \cite{lei2011,salimi2016} with lower values (5-10) assumed only in close proximity of the chromatin, it can be said that the UDG tail should remain mostly extended, during its exploration of the nuclear environment. We also tested the ability to contact a negatively charged sphere of size comparable to the tail length, representing the nucleosome. In these simulations the fluctuating polymer tail, tethered to a mass of 100 units (where each residue has mass 1) representing the UDG core, and a free charged sphere of mass 1000 (the nucleosome), are enclosed in a large box of $100 \times 100 \times 100$ nm. 
In such MD simulations it is typically observed that the RKR motif is able to “find” the negative red sphere, see Figure \ref{fig:figfc1} b), with blue charges positive and red charges negative, within a few ns of simulated time.

\begin{figure}[h!]
\centering
\includegraphics[width=1.1\linewidth]{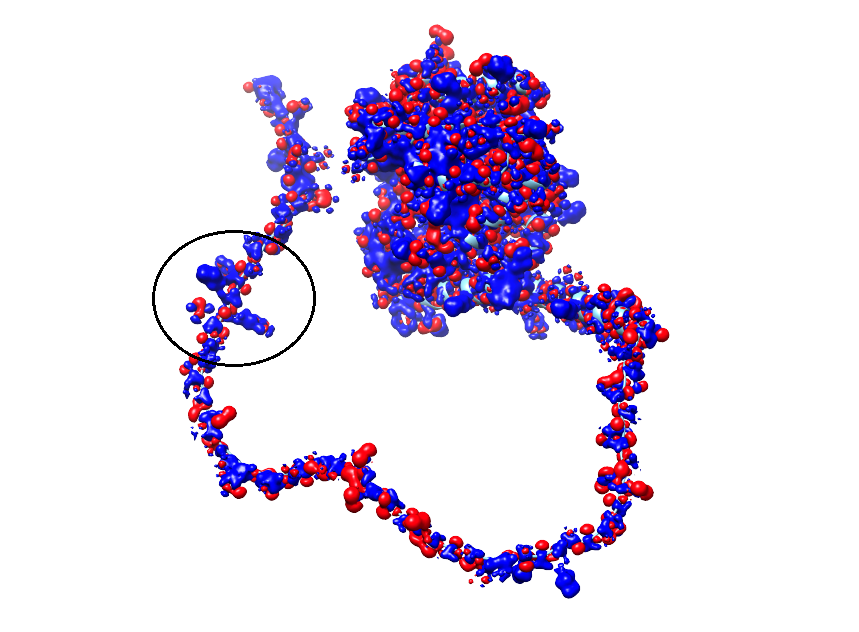}
\caption{Electrostatic potential of complete UDG (hUNG) globular structure and N-terminal tail computed with APBS (for details, see text). The circle encloses the arginine anchor motif in the tail.}
\label{fig:tail_potential_apbs}
\end{figure}

Finally, we have computed the electrostatic potential of the full UDG molecule.
The calculation of electrostatic potentials of proteins is typically based on the linearized Poisson-Boltzmann equation,
\cite{blossey2023} for which a number of publicly available packages exists. We have used both Delphi \cite{li2012,panday2019} and APBS.\cite{jurrus2018} The result shown in Figure \ref{fig:tail_potential_apbs} has been obtained with APBS, using the default parameters of the server. We have also tested other parameters as well as the Delphi implementation with identical results.

In detail, the determination of the electrostatic potential with APBS and its default parameters provided by the web server interface solves 
the linearized Poisson--Boltzmann equation (\texttt{lpbe}) using the automatic multigrid solver (\texttt{mg-auto}) with
single Debye--Hückel boundary conditions (\texttt{bcfl sdh}).
The solute and solvent dielectric constants were set to
$\epsilon_{\mathrm{in}} = 2.0$ and $\epsilon_{\mathrm{out}} = 78.54$,
respectively. A smoothed molecular surface model
(\texttt{srfm smol}) was used with spline-based charge discretization
(\texttt{chgm spl2}), a solvent probe radius of 1.4~\AA{} 
and a smoothing window of 0.3~\AA. The temperature
was set to 298.15~K.
The computational grid was automatically generated and centered on the
biomolecule. The multigrid calculation employed a grid of
$193 \times 193 \times 257$ points, with coarse grid lengths of
$141.63 \times 130.22 \times 191.75$~\AA{} and fine grid lengths of
$103.31 \times 96.60 \times 132.79$~\AA{}, automatically fitted to the
molecular dimensions. This corresponds to a spatial resolution of
approximately $0.50$~\AA{} along the three grid directions.
The electrostatic analysis presented above highlights the presence of localized positively charged regions (shown in blue in \ref{fig:tail_potential_apbs}) within the UDG 
N-terminal tail, notably associated with arginine-rich segments.

\subsection{\label{sec:tail peptide recognition} Tail peptide recognition of the acidic patch: a homology analysis}

While these features are
consistent with a potential interaction with negatively charged surfaces, electrostatics alone does not establish whether this region shares the characteristic sequence signatures of known acidic-patch binding motifs.

To address this, we performed a sequence-based analysis to quantify the similarity between the UDG tail and peptides known to engage the nucleosomal acidic patch. Rather than aiming to detect classical evolutionary homology, this approach evaluates the presence of key sequence and {physico}-chemical features associated with acidic-patch
recognition, such as arginine-anchor motifs, local charge enrichment, and conformational flexibility.\cite{oleinikov2023}
Several viral proteins, chromatin factors, and histone tails interact with the acidic
patch through short basic peptides characterized by a central \emph{arginine anchor}, together with surrounding positively charged and conformationally flexible residues. To capture these structural determinants, four complementary sequence features were combined into a global score. 

To identify motifs most comparable to UDG, particular attention was given to peptides with the highest alignment score, since this term directly measures residue-by-residue similarity to the UDG reference motif. The presence of an \texttt{RKR}-type motif is also an important criterion, as many known nucleosome acidic-patch binders use an arginine residue as an electrostatic ``anchor'' that inserts into the negatively charged pocket formed by histone H2A and H2B residues. Variants of this motif (such as \texttt{KRK}, \texttt{RRK}, or \texttt{RKK}) were considered acceptable when the other parameters remained favorable.

In addition, peptides with an overall positive charge were prioritized, since electrostatic attraction between basic residues (arginine and lysine) and the negatively charged residues of the acidic patch is a well-established feature of nucleosome–protein interactions. Finally, enrichment in serine and proline residues was considered supportive evidence, as such residues are frequently observed in intrinsically disordered regions, which provide the flexibility required for short peptide motifs to adapt to the nucleosome surface.

Together, these criteria allow the identification of peptide motifs that most closely resemble the UDG sequence pattern and are therefore plausible candidates for adopting a similar interaction mode with the nucleosome acidic patch.

The score we developed combines the following factors for which each descriptor captures a different structural property of acidic-patch binding motifs,
with the associated weights chosen to reflect their relative importance in mediating nucleosome recognition:
\begin{enumerate}
\item sequence alignment similarity to the reference UDG motif;
\item the presence of an RKR-like central motif;
\item overall peptide charge;
\item enrichment in serine and proline residues characteristic of intrinsically disordered regions.
\end{enumerate}

\begin{table*}[ht]
\label{tab:motif-similarity}
\centering
\small
\caption{Peptides known to bind to the acidic patch ranked according to the global similarity score defined for UDG for score values $> 0.4$.}
\vspace*{3mm}
\begin{tabular}{|c|c|c|c|c|c|c|}
\hline
\textbf{UniProt} & \textbf{Peptide} & \textbf{Score$_{\mathrm{global}}$} & \textbf{Align$_{\mathrm{norm}}$} & \textbf{Center} & \textbf{Charge} & \textbf{Ser/Pro}
\\
\hline
Q52KI8 & PSPPPARRRRSPSPA & 0.470 & 0.467 & 0.5 & 0.267 & 0.600 \\
E9Q6J5 & LSPSVKRKREVSPPG & 0.460 & 0.200 & 1.0 & 0.200 & 0.400 \\
P26369 & RSRSRDRKRRSRSRD & 0.457 & 0.167 & 1.0 & 0.467 & 0.267 \\
Q8BTI8 & QEPTPAKRKRRSSSS & 0.450 & 0.167 & 1.0 & 0.267 & 0.400 \\
Q569Z6 & RSVSRSRKRRLSSRS & 0.437 & 0.100 & 1.0 & 0.467 & 0.400 \\
Q52KI8 & PSASPPRRRHRPSSP & 0.420 & 0.367 & 0.5 & 0.267 & 0.600 \\
Q8BQ33 & FSHTLSRKRPFRTYT & 0.413 & 0.133 & 1.0 & 0.267 & 0.200 \\
Q52KI8 & RTPSPPPRRRSPSPR & 0.410 & 0.333 & 0.5 & 0.333 & 0.600 \\
Q99M28 & KRDEKERKRRSPSPK & 0.410 & 0.100 & 1.0 & 0.333 & 0.267 \\
Q8R2M2 & TSKSTVRKRQKVAPQ & 0.403 & 0.100 & 1.0 & 0.333 & 0.200 \\
Q6KAQ7 & DSWVSPRKRRLSSSE & 0.403 & 0.100 & 1.0 & 0.133 & 0.400 \\
Q52KI8 & RSPTPPPRRRTPSPP & 0.403 & 0.333 & 0.5 & 0.267 & 0.600 \\
\hline 
\end{tabular}
\end{table*}

The score is constructed in detail as follows:

\subsubsection{Alignment similarity}

The sequence similarity to the reference UDG motif was evaluated over a window of 15 aligned residues. For each aligned position, a score was assigned according to residue similarity according to:
\begin{itemize}
\item identical residue: +2
\item conservative substitution (R$\leftrightarrow$K, D$\leftrightarrow$E, S$\leftrightarrow$T): +1
\item non-conservative substitution: $-1$
\end{itemize}
The alignment score is obtained by summing the contributions across the 15 positions and then normalizing:
\begin{equation}
Score_{\mathrm{align}} =
\frac{\sum_{i=1}^{15} s_i}{2 \times 15}
\end{equation}
where $s_i$ is the substitution score at position $i$.
With this normalization:
\begin{itemize}
\item $Score_{\mathrm{align}} = 1$ corresponds to an identical sequence,
\item values close to $0$ indicate weak similarity,
\item negative values correspond to strongly dissimilar sequences.
\end{itemize}
This term ensures that high scores are obtained only when the peptide resembles the overall motif pattern rather than sharing only a short local feature.
Because reproducing the overall sequence pattern of the UDG tail is the strongest indicator of structural similarity, this descriptor was assigned
the highest weight in the composite score (0.5).

\subsubsection{The central arginine motif}

Because acidic-patch binding motifs are characterized by the presence of a central arginine anchor, the peptide window was examined for the occurrence
of an \texttt{RKR}-type motif or closely related variants. Residues located in the central region of the alignment window (positions 7--11) were therefore analyzed according to the following rule:
\begin{itemize}
\item exact \texttt{RKR} motif $\rightarrow$ score 1
\item close variants (\texttt{KRK}, \texttt{RRK}, \texttt{RKK}, \texttt{KKR}, \texttt{RRR}) $\rightarrow$ score 0.5
\end{itemize}
Because the presence of this basic motif represents a key determinant of
acidic-patch recognition but does not by itself guarantee full motif
similarity, this term was assigned an intermediate weight of 0.3 in the
composite score.

\subsubsection{Charge contribution}

Electrostatic complementarity is a major contributor to acidic-patch recognition.
The net charge of the peptide window was therefore included as
\begin{equation}
Score_{\mathrm{charge}} =
\frac{(\#Arg + \#Lys) - (\#Asp + \#Glu)}{15}
\end{equation}
where the numerator corresponds to the net charge of the 15-residue window. As a high positive charge alone is not sufficient to guarantee acidic-patch
binding, this term was assigned a relatively small weight in the global score (0.1).

\subsubsection{Serine and proline enrichment}

Many acidic-patch interacting peptides originate from intrinsically disordered regions, which frequently exhibit enrichment in serine and proline residues.
These residues promote conformational flexibility and allow the peptide to adapt its structure upon binding.
This contribution was quantified as
\begin{equation}
Score_{\mathrm{SerPro}} =
\frac{(\#Ser + \#Pro)}{15}
\end{equation}
which measures the relative abundance of these residues within the peptide window.
Because this feature supports binding through structural flexibility rather than
direct electrostatic interactions, it was also assigned a modest weight of 0.1.

\subsubsection{Final composite score}

The four contributions were combined into a weighted composite score:
\begin{eqnarray} 
Score_{\mathrm{global}} & = & 
0.5\,Score_{\mathrm{align}}
+0.3\,Score_{\mathrm{center}}\\
&& +0.1\,Score_{\mathrm{charge}} 
+0.1\,Score_{\mathrm{SerPro}} \nonumber
\end{eqnarray}
The weighting scheme prioritizes sequence similarity and the presence of the central arginine motif, while charge and disorder propensity act as secondary
modifiers. This scoring function therefore favors peptides that reproduce the full structural signature of acidic-patch binding motifs rather than sequences
that are merely positively charged.

The peptides listed in Table I
are ranked according to their similarity to the UDG motif. Although the theoretical maximum of the score is 1, such a value would require the simultaneous occurrence of perfect alignment with the UDG sequence, maximal positive charge, and maximal serine/proline enrichment, which is biologically unrealistic. Consequently, the practical range of scores observed in the dataset is significantly lower, and values around 0.45--0.50 correspond to strong similarities detected among the acidic-patch–dependent proteins analyzed here.

\subsection{\label{sec:structure prediction} Structure prediction with AlphaFold}

We finally modeled the binding of the arginine anchor to the H2A/H2B acidic patches on the top and the bottom of the nucleosome through MassiveFold using AlphaFold3, the deep learning tool allowing us to model proteins with DNA.\cite{abramson2024,raouraoua2024}
The modeling of a human nucleosome and the UDG was performed with the sequences of a human nucleosome structure (PDB ID 8VLR) and the UNG sequence P13051-1. AlphaFold3 was used locally with templates to generate 1000 structures (200 seeds * 5 samples). They were then ranked by decreasing order of 0.8ipTM+0.2pTM scores to select the top 10 predictions for the 3D structure representations (Figures 4 and 5).
The {\it template modeling score} or TM-score is a measure of similarity between two protein structures.\cite{zhang2004} The pTM score that derives from it is a predicted TM-score computed by AlphaFold; its definition can be found in the Supplementary Material to \cite{jumper2021},
and the ipTM is a pTM score computed between chains, defined in ref.\cite{evans2021} All renderings were performed with PyMOL 3.1.0.\cite{pyMOL}
\begin{table}[]
\caption{Histone residues in contact within 5 \AA\,  with the RKR motif of the top 10 ranked AlphaFold3 predictions of the UDG stretches (7-30) and their frequence of occurence among them; residues of the acidic patches are identfiable by a cross in the last column.}
\vspace*{3mm}
\begin{tabular}{cccc}
\hline
\multicolumn{1}{|c|}{\textbf{Histone}} & \multicolumn{1}{c|}{\textbf{Residue}} & \multicolumn{1}{c|}{\textbf{\begin{tabular}[c]{@{}c@{}}Number of times found \\ in the 10 top ranked models\end{tabular}}} & \multicolumn{1}{c|}{\textbf{\begin{tabular}[c]{@{}c@{}}Acidic patch\end{tabular}}} \\ \hline
\multicolumn{1}{|c|}{H2A}              & \multicolumn{1}{c|}{TYR 47}           & \multicolumn{1}{c|}{10}                                                                                                    & \multicolumn{1}{c|}{}                             \\
\multicolumn{1}{|c|}{H2A}              & \multicolumn{1}{c|}{GLU 51}           & \multicolumn{1}{c|}{10}                                                                                                    & \multicolumn{1}{c|}{X}                            \\
\multicolumn{1}{|c|}{H2A}              & \multicolumn{1}{c|}{ASP 80}           & \multicolumn{1}{c|}{10}                                                                                                    & \multicolumn{1}{c|}{X}                            \\
\multicolumn{1}{|c|}{H2A}              & \multicolumn{1}{c|}{GLU 82}           & \multicolumn{1}{c|}{10}                                                                                                    & \multicolumn{1}{c|}{X}                            \\
\multicolumn{1}{|c|}{H2B}              & \multicolumn{1}{c|}{GLU 74}           & \multicolumn{1}{c|}{10}                                                                                                    & \multicolumn{1}{c|}{X}                            \\
\multicolumn{1}{|c|}{H2B}              & \multicolumn{1}{c|}{LEU 75}           & \multicolumn{1}{c|}{10}                                                                                                    & \multicolumn{1}{c|}{}                             \\
\multicolumn{1}{|c|}{H2B}              & \multicolumn{1}{c|}{HIS 78}           & \multicolumn{1}{c|}{10}                                                                                                    & \multicolumn{1}{c|}{}                             \\
\multicolumn{1}{|c|}{H2A}              & \multicolumn{1}{c|}{LEU 55}           & \multicolumn{1}{c|}{9}                                                                                                     & \multicolumn{1}{c|}{}                             \\
\multicolumn{1}{|c|}{H2A}              & \multicolumn{1}{c|}{LEU 83}           & \multicolumn{1}{c|}{9}                                                                                                     & \multicolumn{1}{c|}{}                             \\
\multicolumn{1}{|c|}{H2A}              & \multicolumn{1}{c|}{GLU 54}           & \multicolumn{1}{c|}{7}                                                                                                     & \multicolumn{1}{c|}{X}                            \\
\multicolumn{1}{|c|}{H2A}              & \multicolumn{1}{c|}{ASN 58}           & \multicolumn{1}{c|}{5}                                                                                                     & \multicolumn{1}{c|}{}                             \\
\multicolumn{1}{|c|}{H2A}              & \multicolumn{1}{c|}{ALA 50}           & \multicolumn{1}{c|}{2}                                                                                                     & \multicolumn{1}{c|}{}                             \\
\multicolumn{1}{|c|}{H2B}              & \multicolumn{1}{c|}{VAL 17}           & \multicolumn{1}{c|}{2}                                                                                                     & \multicolumn{1}{c|}{}                             \\
\multicolumn{1}{|c|}{H2A}              & \multicolumn{1}{c|}{GLU 81}           & \multicolumn{1}{c|}{1}                                                                                                     & \multicolumn{1}{c|}{X}                            \\
\multicolumn{1}{|c|}{H2B}              & \multicolumn{1}{c|}{SER 81}           & \multicolumn{1}{c|}{1}                                                                                                     & \multicolumn{1}{c|}{}                             \\
\multicolumn{1}{|c|}{H2B}              & \multicolumn{1}{c|}{GLU 82}           & \multicolumn{1}{c|}{1}                                                                                                     & \multicolumn{1}{c|}{X}                            \\
\multicolumn{1}{|c|}{H2B}              & \multicolumn{1}{c|}{GLN 16}           & \multicolumn{1}{c|}{1}                                                                                                     & \multicolumn{1}{c|}{}                             \\
\multicolumn{1}{|c|}{H2A}              & \multicolumn{1}{c|}{ASN 79}           & \multicolumn{1}{c|}{1}                                                                                                     & \multicolumn{1}{c|}{}                             \\
\hline
\end{tabular}
\end{table}

The nucleosome alone is well modeled with a RMSD (Root Mean Square Deviation) = 0.92 Ångströms and TM-score = 0.989 when aligning the best ranked AlphaFold3 model to the 8VLR structure. This model is depicted in Figure 4 (Top), with the acidic patches of the H2A and H2B highlighted as red surfaces.\cite{oleinikov2023} When including a 7-30 stretch of UDG centered on the RKR motif in the modeling, AlphaFold3 systematically binds this arginine anchor stretch to the acidic patches, as shown with the top 10 ranked predictions in Figure 4 (Bottom). The arginines and the lysine of the stretch are highlighted as blue spheres. As listed in Table 2, the RKR motif is systematically predicted binding to several residues of the acidic patches in the top 10 models, namely E51, D80, E82 of histone H2A and E74 of histone H2B. E54 of histone H2A is also often identified in contact, being found 7 times in contact with the arginine anchor motif. Two other residues of the acidic patch, E81 of H2A and E82 of H2B are also found in contact to a lesser extent. The acidic patches are clearly a favored position for the UDG stretches, with the RKR motif in particular binding to several residues of the patch, however, the orientation of the peptides in the vicinity is not exactly the same, offering a subtle clue on the fact that this binding is favored but not too strong to allow some movements of the UDG peptide.

\begin{figure}[t!]
\centering
\includegraphics[width=.85\linewidth]{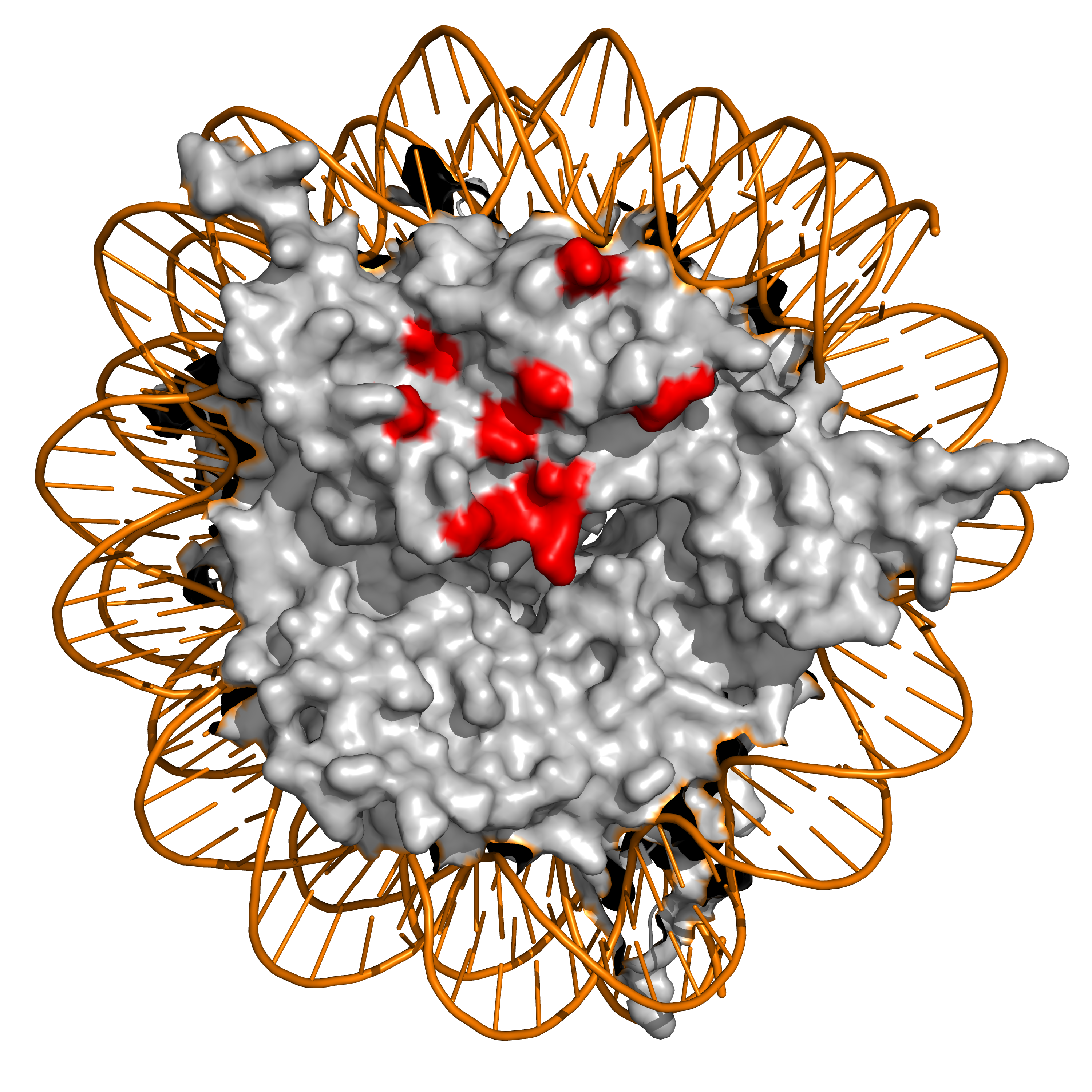}
\includegraphics[width=.85\linewidth]{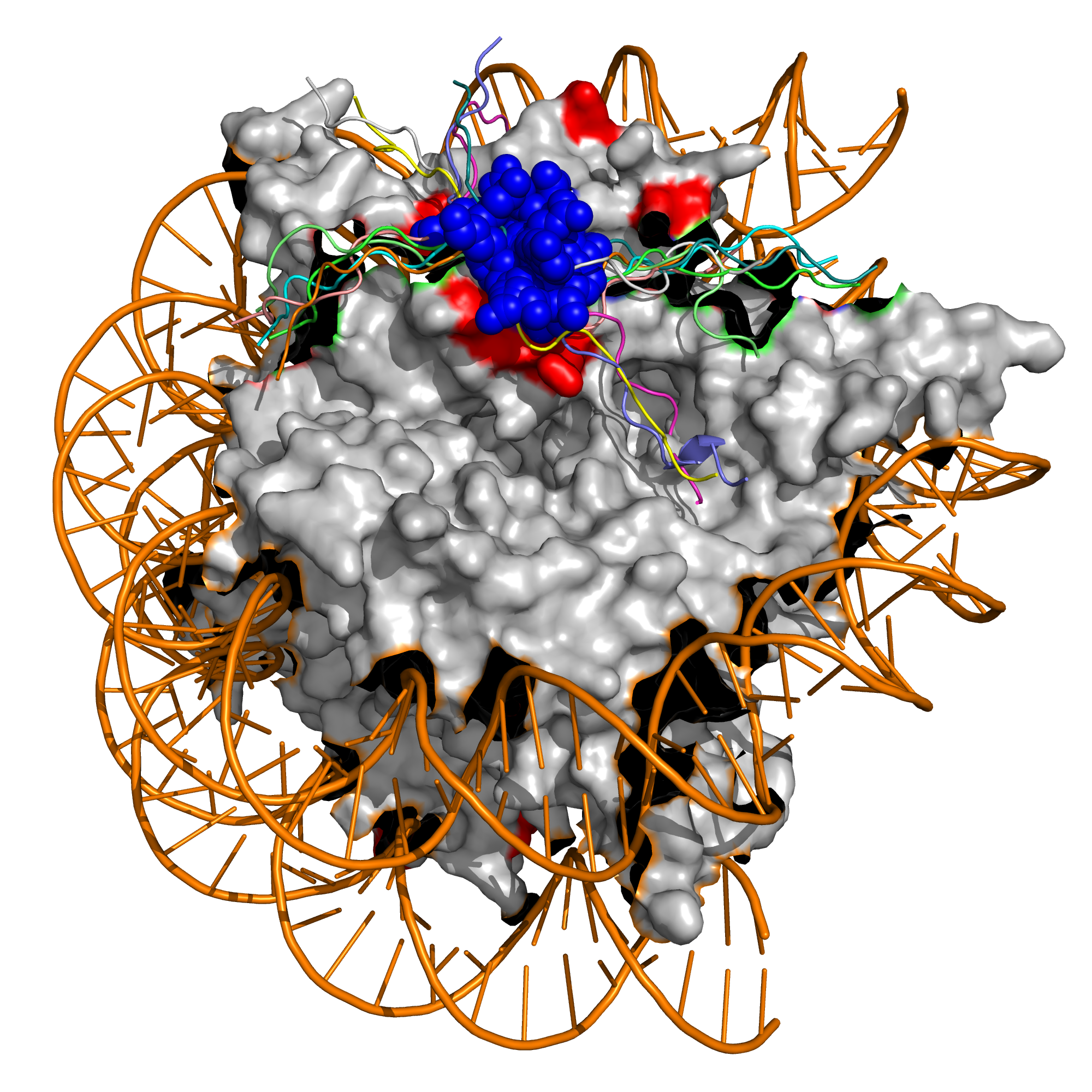}
\caption{Tail peptides binding to the acidic patches. The top structure is a top view of a nucleosome (from PDB 8VLR) where the DNA is represented as cartoon in orange and the histones core surface is shown in gray; the acidic patches of both H2A and both H2B are colored in red; only the core of the histones are depicted, with only a small part of H2A and H3 terminal tails extending over the DNA on top left and right respectively, like in the 8VLR structure. The bottom structure shows the systematic binding of UDG peptides (residues 7 to 30) to the H2A/H2B acidic patches in the top 10 AlphaFold3 predictions ranked by decreasing order of 0.8ipTM+0.2pTM; the arginines and lysine of the UDG anchors depicted as blue spheres bind to the acidic patches; the remaining of the peptides is shown in cartoon mode in various colours.}
\label{fig:peptides}
\end{figure}

To further evaluate the propensity of the arginine anchor to bind to the acidic patches, we modeled the full UDG with the nucleosome (Figure 5). The top 10 ranked AlphaFold3 predictions always show the core of the UDG binding to an easily accessible section of the DNA, while the long N-terminal tail reaches the H2A/H2B acidic patches with its arginine anchor in seven cases, on the top H2A/H2B or on the bottom ones. Here, the flexible UDG N-terminal tail is long enough to perform such a binding, which could help the UDG to stabilize itself to its targeted thymine. Like the modeling with UDG (7-30) stretches, the E51 and D80 of histone H2A make contacts with the UDG anchor in all the seven models, while contacts with E82 of H2A are found five times, with E74 of H2B four times and with E54 of H2A one time, highlighting the importance of these key residues in the binding of the UDG to the nucleosome. This binding being favored, it does not look too strong, with a few of the models showing a N-terminal tail not binding to the histones, suggesting an optional binding of the anchor and possible movements of the UDG around the nucleosome.

\begin{figure}[t!]
\includegraphics[width=0.95\linewidth]{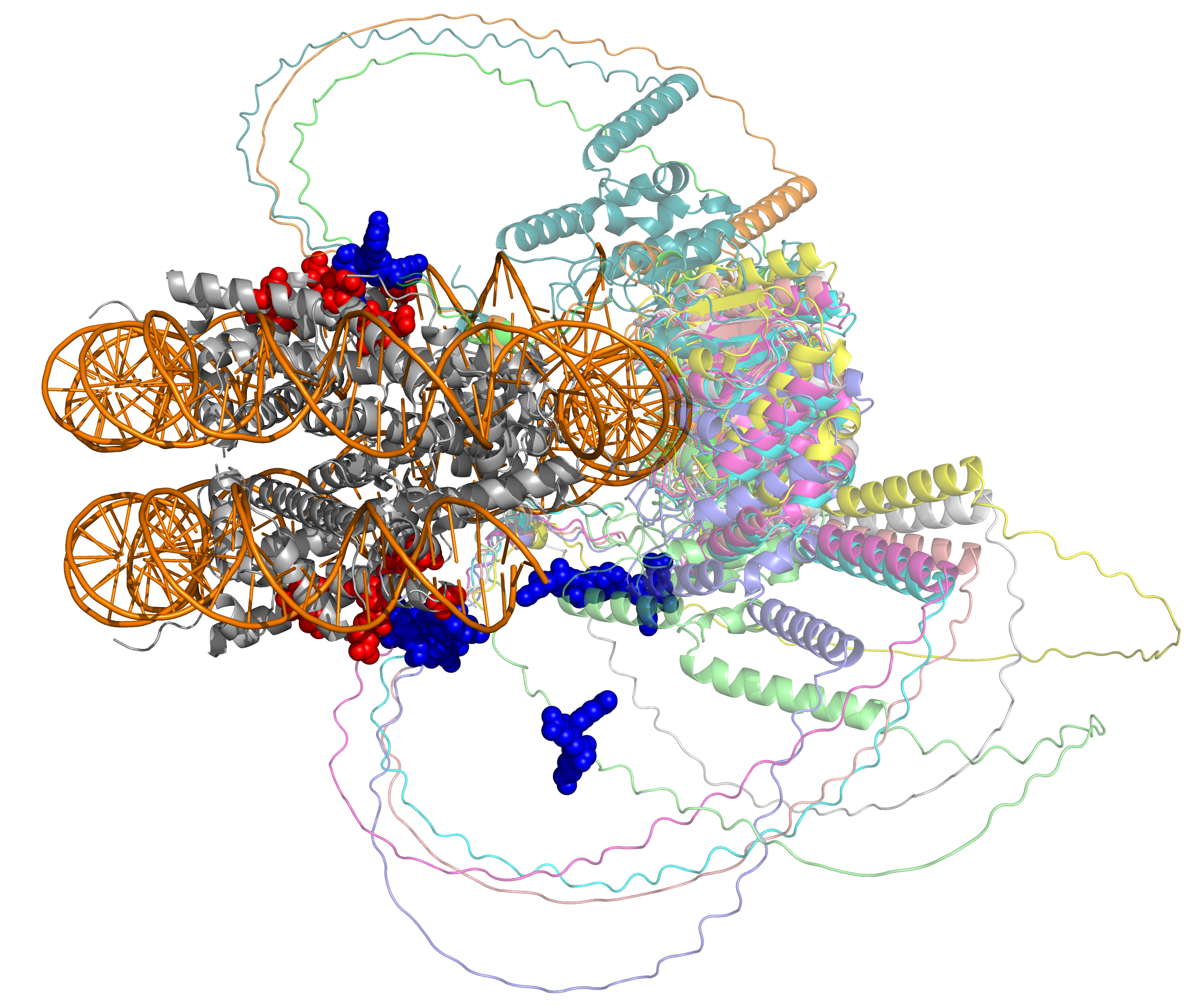}
\caption{Binding of full UDG (in light transparency) to the nucleosome (from PDB 8VLR), all in cartoon representation. In this structure, the arginine anchor binds to the acidic patches at the bottom or top of the nucleosome, while the
active site in the globular domain binds to a thymine nucleotide in the DNA gyre in an easily accessible region of the nucleosome. Top 10 AlphaFold3 predictions based on 0.8iPTM+0.2pTM were selected for representation. The acidic patches on both H2A and both H2B are represented in red spheres while the arginine anchors on the UDG N-terminal tails are represented by blue spheres.}
\label{fig:tail}
\end{figure}

\section{\label{sec:Conclusions} Conclusions and Discussion}

In this work we have established the following elements for the interactions of the DNA repair enzyme UDG with nucleosomes. Firstly, the N-terminal tail of the enzyme has a charge distribution carrying  a motif that not only serves a nuclear localization signal (NLS), but also as an arginine anchor. The electrostatic properties of the tail were studied by MD simulations of a simple coarse-grained polymer model and the calculation of the electrostatic potential via the Poisson-Boltzmann equation. Subsequently we have shown, via a homology analysis, that the NLS  motif also shares sequence similarity with known sequences of transcription factors that use the acidic patch of the nucleosome as their binding
platform. Finally, a structural analysis performed with AlphaFold 
shows that the UDG tail motif is predicted to bind to the acidic patches on the nucleosome. The final build of a structure for the nucleosome in complex with the full UDG complex, i.e. its globular domain complete with the tail, demonstrates that the tail motif either binds to the top or bottom acidic patch on the nucleosome. 

These results lead us to the following view of the role of the
UDG tail in the interaction of the enzyme with the nucleosome. 
The nucleosome is a compact structure which impedes 
the diffusion of the enzyme along DNA as well as its access to
defects: we have analyzed this situation in earlier work via
MD simulations. The tail charge motif act as an arginine anchor to
the acidic patch of the nucleosome and therefore increases the probability of its presence right on this structure. The binding of the tail motif is found to be robust and it appears flexible enough to allow UDG to sample the nucleosome. Combined with
its tail length being seemingly nicely adapted to the size of the nucleosome, this mechanism allows the enzyme to map out the nucleosomal surface for sequence defects.

It will be interesting, e.g., to validate the role of the tail in future experiments with kinetic assays. Such assays have recently proven that more biologically realistic DNA sequences than the strongly positioning Widom 601 sequence enhance the efficiency of UDG on nucleosomes.\cite{barbosa2025} Our
finding may render the possibilities for the interaction of 
UDG with nucleosomes yet more dynamic than previously thought.
It is thus another example for the subtle role that electrostatic interactions can play in biological systems.
\\

\begin{acknowledgments}
This article is dedicated to the memory of Rudi Podgornik, who
has been an inspiring collaborator of the lead author of this work. 
DNA and protein electrostatics has been one Rudi's many interests in biophysics, and we hope he would have liked the 
little twist of the role of electrostatics in protein-DNA 
interactions we have presented here. This work has been supported by the ANR grant “Dyprosome” (ANR-21-CE45-0032). 
\end{acknowledgments}

\section*{Data Availability Statement}

The data that support the findings of this study are available within the article.

\section*{Author Declarations}

{\bf Conflict of Interest}
\\

The authors have no conflicts of interest to disclose.
\\

{\bf Author contributions}
\\

{\bf Safwen Ghediri}: Investigation (equal); Writing (supporting).
{\bf Guillaume Brysbaert}: Investigation (equal); Writing (supporting). 
{\bf Fabrizio Cleri}: Investigation (equal); Funding acquisition
(lead); Writing (supporting).
{\bf Ralf Blossey}: Conceptualization (lead); Investigation (lead); Supervision (lead); Writing (lead); Funding acquisition (supporting); 

\bibliography{references}

\end{document}